\documentclass[onecolumn,superscriptaddress,nofootinbib,longbibliography,preprintnumbers]{revtex4-2}
\usepackage{amsmath,amsthm}
\usepackage{amssymb,amsfonts}
\usepackage{bm}
\usepackage[colorlinks=true, linkcolor=blue, citecolor=blue, urlcolor=blue]{hyperref}
\usepackage{graphicx}% Include figure files
\usepackage{float}
\usepackage{braket}
\usepackage{subfigure}
\usepackage{cases}
\usepackage{multirow}
\numberwithin{equation}{section}

\newcommand{\diff}{\mathrm{d}}

\newcommand{\Tr}{\mathrm{Tr}}
\newcommand{\tTr}{\mathrm{tTr}}

\newcommand{\e}{\mathrm{e}}

\usepackage{tikz,xcolor}
\definecolor{lime}{HTML}{A6CE39}
\DeclareRobustCommand{\orcidicon}{%
	\begin{tikzpicture}
	\draw[lime, fill=lime] (0,0) 
	circle [radius=0.16] 
	node[white] {{\fontfamily{qag}\selectfont \tiny ID}};	\draw[white, fill=white] (-0.0625,0.095) 
	circle [radius=0.007];	\end{tikzpicture}
	\hspace{-2mm}}
\foreach \x in {A,...,Z}{%
	\expandafter\xdef\csname orcid\x\endcsname{\noexpand\href{https://orcid.org/\csname orcidauthor\x\endcsname}{\noexpand\orcidicon}}
}

\newcommand{\FNAL}{\affiliation{
Fermi National Accelerator Laboratory, Batavia, Illinois, USA}}

\newcommand{\JLAB}{\affiliation{
Thomas Jefferson National Accelerator Facility, Newport News, Virginia 23606, USA}}

\newcommand{\TU}{\affiliation{Center for Computational Sciences, University of Tsukuba, Tsukuba, Ibaraki 305-8577, Japan}}

\newcommand{\UTokyo}{\affiliation{Graduate School of Science, The University of Tokyo, Bunkyo-ku, Tokyo, 113-0033, Japan}}

\begin{document}

\title{$SU(2)$ principal chiral model with tensor renormalization group on a cubic lattice}

\author{Shinichiro Akiyama~\orcidA{}\,}
\email{akiyama@ccs.tsukuba.ac.jp}
\TU
\UTokyo

\author{Raghav G. Jha~\orcidB{}\,}
\email{raghav.govind.jha@gmail.com}
\JLAB

\author{Judah Unmuth-Yockey~\orcidC{}\,}
\email{jfunmuthyockey@gmail.com}
\FNAL

\preprint{JLAB-THY-24-4047,UTCCS-P-154,FERMILAB-PUB-24-0308-T}

\begin{abstract}
  We study the continuous phase transition and thermodynamic observables in the three-dimensional Euclidean $SU(2)$ principal chiral field model with the triad tensor renormalization group (tTRG) and the anisotropic tensor renormalization group (ATRG) methods. Using these methods, we find results that are consistent with previous Monte Carlo estimates and the predicted renormalization group scaling of the magnetization close to criticality. These results bring us one step closer to studying finite-density QCD in four dimensions using tensor network methods. 
\end{abstract}

\maketitle

\onecolumngrid

\section{Introduction}

%%%%%%%%%%%%%%%%%%%
In recent decades, significant progress has been made in
understanding the non-perturbative physics of quantum chromodynamics
(QCD) through lattice gauge theory based on Wilson's
insights. However, a major challenge in the Monte Carlo (MC) simulations arises when attempting to compute real-time dynamics or simulate at finite-baryon density (chemical potential) or in the presence of topological theta term. This obstacle is primarily due to a signal-to-noise issue called the ``sign problem'', where relevant gauge-invariant quantities are exponentially suppressed relative to their error margins. This hinders a comprehensive understanding of the QCD phase diagram at finite baryon density and temperature which is crucial for a proper understanding of a wide range of rich phenomena. For a recent comprehensive review, we refer the reader to  Refs.~\cite{deForcrand:2009zkb,Nagata:2021ugx}. 

Given the limitations of existing sampling-based numerical tools, there is a growing interest in exploring alternative methods ranging from classical tensor networks to quantum computing. Tensor network techniques, particularly when integrated into a blocking algorithm known as the
``tensor renormalization group'' (TRG)~\cite{Levin:2006jai}, have shown promise in addressing these limitations. 
While TRG methods have been successfully applied to
several spin systems and lattice gauge theories~\cite{Shimizu:2014uva,Shimizu:2014fsa,Dittrich:2014mxa,Shimizu:2017onf,Unmuth-Yockey:2018ugm,Kuramashi:2018mmi,Unmuth-Yockey:2018xak,Bazavov:2019qih,Kuramashi:2019cgs,Butt:2019uul, Jha:2020oik,Fukuma:2021cni, Hirasawa:2021qvh,Milde:2021vln,Kuwahara:2022ubg,Akiyama:2022eip,Bloch:2022vqz,Akiyama:2023hvt,Yosprakob:2023tyr,Asaduzzaman:2023pyz}, efforts to employ them in studying models in more than two Euclidean dimensions have still been difficult. This difficulty was addressed by TRG algorithms which scale better in a larger number of dimensions~\cite{Adachi:2019paf,Oba:2019csk,Kadoh:2019kqk,Nakayama:2023ytr} compared to the higher-order TRG (HOTRG), a variant of the TRG method, which scales like $O(D^{4d-1})$ in $d$-dimensional Euclidean systems with bond dimension $D$~\cite{Xie:2012mjn}.   

Two prominent approaches are the triad TRG (tTRG)~\cite{Kadoh:2019kqk}, and the anisotropic TRG (ATRG)~\cite{Adachi:2019paf}, which are the main focus of this work. Both these algorithms have improved scaling in computational time and memory relative to the HOTRG algorithm, with the computational time in the tTRG scaling like $O(D^{d+3})$, and the ATRG scaling like $O(D^{2d+1})$. These two TRG algorithms have been applied for various higher-dimensional lattice theories~\cite{Akiyama:2020ntf,Akiyama:2020soe,Akiyama:2021zhf,Bloch:2021mjw,Bloch:2021uup,Jha:2022pgy,Kuwahara:2022ubg,Akiyama:2022eip,Akiyama:2023hvt}.
However, even with the improved time complexity, the two methods make additional approximations in different ways leading to them potentially discarding different physics during coarse-graining. Because of these differences, it is important to quantify their respective efficacies to understand which approximations are superior, if at all.

A further difficulty is the computational cost associated with tensor formulations of actions possessing non-Abelian symmetries~\cite{Liu:2013nsa,Meurice:2020pxc,Yosprakob:2023jgl}.  These non-Abelian symmetries prompt the existence of additional quantum numbers, giving rise to more complicated constraints between degrees of freedom.  The transition from Abelian symmetries, where simple conservation-law constraints appear, to non-Abelian symmetries where milder
constraints manifest, \emph{e.g.} triangle inequalities, drastically increases the memory cost of tensor networks.  Because of this, the algorithmic improvements provided by the tTRG and the ATRG are encouraging, but it is still unclear how these algorithms handle these constraints under blocking.

To understand the potential of these algorithms as a means to eventually probe QCD, it is often useful to consider simpler models with similar properties. Notably, it has been suggested that the chiral phase transition
in the two massless flavors approximation of QCD shares critical
exponents with the three-dimensional $SU(2)$ principal chiral model
($O(4)$ nonlinear sigma model)~\cite{PhysRevD.29.338,Wilczek:1992sf,Rajagopal:1992qz,Engels:2005rr,Ejiri:2009ac}. 
Leveraging this connection,
investigations have been conducted using MC and other
numerical approaches to shed light on this model~\cite{Kanaya:1994qe,PhysRevD.55.362,Hasenbusch_2001,Engels:2003nq,Engels:2011km,Xu:2011sj,Luo:2022eje}.

In this study, we focus on exploring the three-dimensional $SU(2)$ principal chiral
model using tensor network methods. 
We construct the tensor network representation for the path integral based on the character expansion, which allows us to preserve the original $SU(2)$ symmetry even after the finite truncation of the irreducible representations.
Since the symmetry is preserved, it is naturally expected that we can obtain the correct scaling behavior of the $O(4)$ model. We note that this kind of truncation effect has been recently investigated in the two-dimensional $O(2)$ model with tensor network methods~\cite{Zhang:2021dnz} and our work is the first tensorial attempt in three dimensions extending the preliminary work by authors in Ref.~\cite{Akiyama:2023fyk}. By employing the ATRG and tTRG algorithms, the goal is to compute canonical quantities such as internal
energy, magnetization, and the critical coupling associated with the continuous phase transition. We also comment on the critical exponents and the fixed-point behavior of the algorithms used in this paper.

\section{The model}
\label{sec:model}
The $SU(2)$ principal chiral model in three dimensions has a continuum action given by 
\begin{align}
	S
	=
	\frac{J}{4}
	\int\diff^{3}x
	\Tr\left[
	\sum_{\nu=1}^{3}\partial_{\nu}U(x)^{\dag}\partial_{\nu}U(x)
	\right],
 \label{eq:Act1} 
\end{align}
where $J$ is the coupling constant and 
$U(x)$ are elements of $SU(2)$. The action in \eqref{eq:Act1} is invariant under an $SU(2)_{\rm{L}} \otimes SU(2)_{\rm{R}}$ global symmetry which acts on $U$ and $U^{\dagger}$ as $U(x) \to G_{\rm{L}} U^{\prime}(x) G_{\rm{R}}^{\dagger}$ and 
$U^{\dagger}(x) \to G_{\rm{R}} U^{\prime \dagger}(x) G_{\rm{L}}^{\dagger}$ where $G$ is an element of the $SU(2)$. This model is equivalent to the $O(4)$ nonlinear sigma model because of the well-known homomorphism i.e., $SU(2) \otimes SU(2) \equiv SO(4)$ and was first considered by Polyakov as a toy model to understand the chiral symmetry breaking in QCD \cite{Polyakov:1987ez}. 
The Euclidean lattice action is: 
\begin{align}
\label{eq:action}
 	S = -\frac{J}{2}
	\sum_{n,\nu}
	\Tr\left[
	U(n)U(n+\hat{\nu})^{\dag}
	\right]
	,
\end{align}
where the $SU(2)$ group elements live on the sites, $n$, of the lattice, and the $\nu$-sum is over the three orthogonal directions of the cubic lattice.
The path integral on the lattice is given by
\begin{align}
\label{eq:path_int}
	Z= \int
	\prod_{n}\diff U(n)
	~\e^{-S}
	,
\end{align}
and the group integration is the typical Haar integral.
We assume periodic boundary conditions for all directions.

The average internal energy density can be defined by a derivative of the thermodynamic potential $f(J) \equiv -\log{Z(J)}$,
\begin{align}
\label{eq:avg-act}
	\braket{e}
	=
	\frac{1}{V}\frac{\partial}{\partial J} f(J)
	.
\end{align}
One can explicitly break the global symmetry by including an external field term in the action,
\begin{align}
    S_{H} = - H \sum_{n} \Tr[U(n)].
\end{align}
The average value of this on-site interaction term can also be defined using a derivative of $f(J, H)$,
\begin{align}
\label{eq:mag}
    \braket{m} \equiv \frac{1}{V} \left\langle \sum_{n} \Tr[U(n)] \right\rangle = -\frac{1}{V} \frac{\partial}{\partial H} f(J, H).
\end{align}
In the following section we will provide tensor network descriptions of the path integral and these expectation values, along with one more useful quantity.

\subsection{Tensor network construction}
\begin{figure}
    \centering
    \includegraphics[width=16cm]{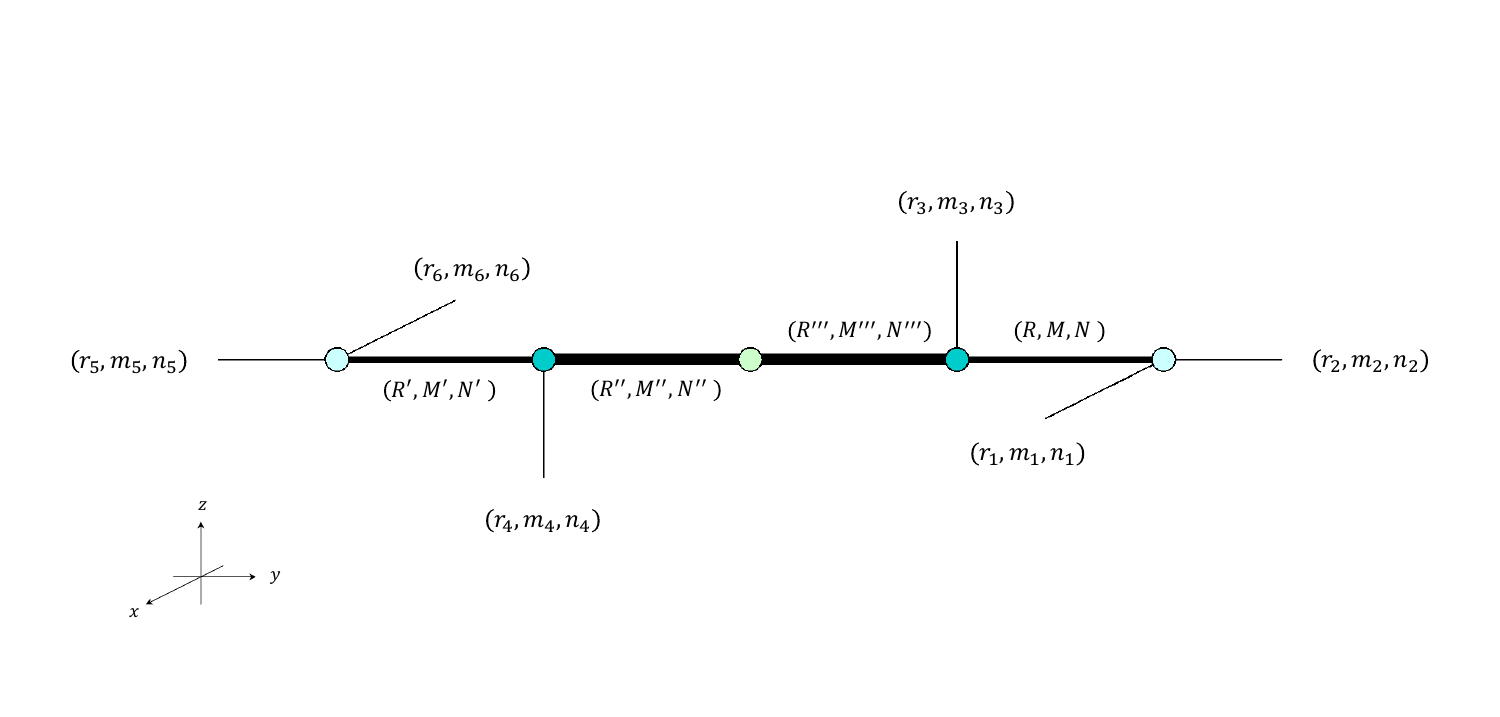}
    \caption{
    Tensor network diagram of Eq.~\eqref{eq:def_t}.
    Each three-leg tensor denotes the Clebsch-Gordon coefficients and the two-leg tensor at the center represents Kronecker delta function.
    Eq.~\eqref{eq:T-with-h} is also denoted by the same diagram but with Eq.~\eqref{eq:hmatrix} as the matrix at the center.
    Due to the fusion rule, internal lines have larger bond dimensions than external ones, as shown by their thickness.
    }
    \label{fig:fundamental_tensor}
\end{figure}

Tensor network constructions for this model without $S_{H}$ have been considered before in Refs.~\cite{Liu:2013nsa,Meurice:2020pxc,Luo:2022eje}.  Here we briefly review the construction using the character expansion method, as well as provide a construction in the case of an external field.
\subsubsection{Basic construction}

Our goal is to represent Eq.~\eqref{eq:path_int} as a tensor network such that
\begin{align}
\label{eq:tn_z}
	Z
	=
	\tTr\left[
	\prod_{n}T_{n}
	\right]
	,
\end{align}
where $\tTr$ is the tensor trace over all indices,
and $T_{n}$ is a tensor at site $n$ whose indices are suppressed.  
Using the character expansion \cite{Drouffe:1983fv, Campostrini:1994sh}, we have
\begin{align}
\label{eq:original_ce}
	\exp\left[
	\frac{J}{2}
	\Tr[U(n)U(n+\hat{\nu})^{\dag}]
	\right]
	=
	\sum_{r}
	F_{r}(J) \chi_{r}(U(n)U(n+\hat{\nu})^{\dag})
	,
\end{align}
with the coefficients written in terms of the modified Bessel function of the first kind as: 
\begin{align}
\label{eq:Fr}
	F_{r}(J)
	=
	\frac{2(2r+1)}{J}I_{2r+1}(J)
	.
\end{align}
Since the character $\chi_{r}(U)$ is the trace of the matrix representation of the group element $U$ in the irreducible representation of $r$, we have,
% Page 89, Eq (19) Varshalovich et al.
\begin{align}
	\chi_{r}(U(n)U(n+\hat{\nu})^{\dag})
	=
	\sum_{m,k}
	D^{(r)}_{mk}(U(n))D^{(r)*}_{mk}(U(n+\hat{\nu}))
	.
\end{align}
This allows each group element to be integrated individually, generating constraints associated with the sites of the lattice involving Clebsch-Gordan coefficients.
After completing all the integrals,
a local tensor at a site $n$ is given by
\begin{align}
\label{eq:def_t}
	T_{n;
    x y z z' y' x'
	}
	&=
	\sqrt{\prod_{p=1}^{6}F_{r_{p}}(J)}
	%\times
	\sum_{R=|r_{1}-r_{2}|}^{r_{1}+r_{2}}
	\sum_{R'''=|R-r_{3}|}^{R+r_{3}}
	\sum_{R''=|r_{4}-R'|}^{r_{4}+R'}
	\sum_{R'=|r_{5}-r_{6}|}^{r_{5}+r_{6}}
	\sum_{M,N}
	\sum_{M''',N'''}
	\sum_{M'',N''}
	\sum_{M',N'}
	\nonumber\\
	&\times
	C^{RM}_{r_{1}m_{1}r_{2}m_{2}}
	C^{RN}_{r_{1}n_{1}r_{2}n_{2}}
	C^{R'''M'''}_{RMr_{3}m_{3}}
	C^{R'''N'''}_{RNr_{3}n_{3}}
	C^{R''M''}_{r_{4}m_{4}R'M'}
	C^{R''N''}_{r_{4}n_{4}R'N'}
	C^{R'M'}_{r_{5}m_{5}r_{6}m_{6}}
	C^{R'N'}_{r_{5}n_{5}r_{6}n_{6}}
	\nonumber\\
	&\times
	\frac{1}{2R'''+1}
	\delta_{R''',R''}
	\delta_{M''',M''}
	\delta_{N''',N''}
	,
\end{align}
where $C^{j m}_{j_1 m_1 j_2 m_2}$ is the Clebsch-Gordan coefficient
and $x \equiv (r_{1} m_{1} n_{1})$ is the collective index notation used often going forward.  This tensor can be understood in terms of a sequence of smaller tensor contractions---whose nonzero elements are given by the Clebsch-Gordan coefficients---which are illustrated in Fig.~\ref{fig:fundamental_tensor}.  The tensor in Eq.~\eqref{eq:def_t} can be inserted into Eq.~\eqref{eq:tn_z} for an expression of the path integral using a tensor network contraction.

The field regularization is achieved by introducing a cutoff parameter $r_{\rm max}$ in Eq.~\eqref{eq:original_ce} as
\begin{align}
	\exp\left[
	\frac{J}{2}
	\Tr[U(n)U(n+\hat{\nu})^{\dag}]
	\right]
	&\simeq
	\sum_{r=0}^{r_{\rm max}}
	F_{r}(J)\chi_{r}(U(n)U(n+\hat{\nu})^{\dag})
	,
\end{align}
which preserves the global symmetry and does not affect the integration.  Using identical methods, we can express the expectation value given in Eq.~\eqref{eq:avg-act} as a ratio of two tensor network contractions.

\subsubsection{Impurity tensor for average action}

Using the expressions in Eqs.~\eqref{eq:avg-act} and \eqref{eq:original_ce}, the derivative of $f(J)$ affects the character expansion in the following way:
\begin{align}
	\frac{\partial}{\partial J}
	\exp\left[
	\frac{ J}{2}
	\Tr[U(n)U(n+\hat{\nu})^{\dag}]
	\right]
	&\simeq
	\sum_{r=0}^{r_{\rm max}}
	\frac{\partial F_{r}( J)}{\partial J}
	\chi_{r}(U(n)U(n+\hat{\nu})^{\dag})
	\nonumber\\
	&=
	\sum_{r=0}^{r_{\rm max}}
	\widetilde {F}_{r}( J)
	\chi_{r}(U(n)U(n+\hat{\nu})^{\dag})
	,
\end{align}
with
\begin{align}
\label{eq:fwidetilde }
	\widetilde {F}_{r}( J)
	=
	\frac{2(2r+1)}{ J^{2}}\left[
	2rI_{2r+1}( J)+ J I_{2r+2}( J)
	\right]
	.
\end{align}
This minor modification gives an expression for
$\braket{e}$ in terms of a tensor network that includes two adjacent `impurities',
\begin{align}
\label{eq:atrg_s_imp}
	\braket{e}
	=
	-\frac{3}{Z}
	\tTr\left[
	R_{n+\hat{\nu}}S_{n}\prod_{n'\neq n,n+\hat{\nu}}T_{n'}
	\right]
	,
\end{align}
and is the ratio of two scalars, tensor network contractions.
This is possible because of the translation invariance of our lattice. The
impurities are given by
\begin{align}
\label{eq:Rimp}
	&R_{n+\hat{\nu};xyzz'y'x'}
	=
	\sqrt{\frac{\widetilde {F}_{r_{\nu'}}( J)}{F_{r_{\nu'}}( J)}}
	T_{n+\hat{\nu};xyzz'y'x'}
	,
\end{align}
and
\begin{align}
\label{eq:Simp}
	&S_{n;xyzz'y'x'}
	=
	\sqrt{\frac{\widetilde  {F}_{r_{\nu}}( J)}{F_{r_{\nu}}( J)}}
	T_{n;xyzz'y'x'} .
\end{align}
Note that the $\nu$-directional coarse-graining should be carried out first. We will now discuss the construction of a tensor network when the model includes an external field term in the action.

\subsubsection{Pure and Impure tensors for magnetization}

Consider the action now with an external field term,
\begin{align}
    S = -\frac{ J}{2}
	\sum_{n,\nu}
	\Tr\left[
	U(n)U(n+\hat{\nu})^{\dag}
	\right]
    -
    H \sum_{n} 
    \Tr\left[U(n) \right]
.
\end{align}
In this case, we need to modify the pure tensor, because we have to deal with the finite magnetic field, $H$, which gives us a new exponential factor.  This factor can be expanded as before, giving
\begin{align}
	\e^{H \Tr[U(n)]}
	=
	\sum_{r_{H}}
	F_{r_{H}}(2H)\chi_{r_{H}}(U(n))
	=
	\sum_{r_{H}}
	F_{r_{H}}(2H)
	\sum_{m_{H}=-r_{H}}^{r_{H}}
	D^{(r_{H})}_{m_{H}m_{H}}(U(n))
	,
\end{align}
where $F_{r_{H}}$ is defined as in Eq.~\eqref{eq:Fr}.
We can use the character expansion to isolate the individual group elements and perform the Haar integration over each one.  The integrals from before, in the absence of an external field, are modified but straightforward.  The resulting local tensor at site $n$ is given by,

\begin{align}
\label{eq:T-with-h}
	&
	T_{n;
	xyzz'y'x'
	}
	=
	\sqrt{\prod_{p=1}^{6}F_{r_{p}}( J)}
	\sum_{R=|r_{1}-r_{2}|}^{r_{1}+r_{2}}
	\sum_{R'''=|R-r_{3}|}^{R+r_{3}}
	\sum_{R''=|r_{4}-R'|}^{r_{4}+R'}
	\sum_{R'=|r_{5}-r_{6}|}^{r_{5}+r_{6}}
	\sum_{M,N}
	\sum_{M''',N'''}
	\sum_{M'',N''}
	\sum_{M',N'}
	\nonumber\\
	&\times
	C^{RM}_{r_{1}m_{1}r_{2}m_{2}}
	C^{RN}_{r_{1}n_{1}r_{2}n_{2}}
	C^{R'''M'''}_{RMr_{3}m_{3}}
	C^{R'''N'''}_{RNr_{3}n_{3}}
	C^{R''M''}_{r_{4}m_{4}R'M'}
	C^{R''N''}_{r_{4}n_{4}R'N'}
	C^{R'M'}_{r_{5}m_{5}r_{6}m_{6}}
	C^{R'N'}_{r_{5}n_{5}r_{6}n_{6}}
	\nonumber\\
	&\times
	\frac{1}{2R''+1}
	\sum_{r_{H}}
	F_{r_{H}}(2 H)
	\sum_{m_{H}=-r_{H}}^{r_{H}}
	C^{R''M''}_{R'''M'''r_{H}m_{H}}
	C^{R''N''}_{R'''N'''r_{H}m_{H}}
	\nonumber\\
	&=
	\sqrt{\prod_{p=1}^{6}F_{r_{p}}( J)}
	%\nonumber\\
	%&\times
	\sum_{R=|r_{1}-r_{2}|}^{r_{1}+r_{2}}
	\sum_{R'''=|R-r_{3}|}^{R+r_{3}}
	\sum_{R''=|r_{4}-R'|}^{r_{4}+R'}
	\sum_{R'=|r_{5}-r_{6}|}^{r_{5}+r_{6}}
	\sum_{M,N}
	\sum_{M''',N'''}
	\sum_{M'',N''}
	\sum_{M',N'}
	\nonumber\\
	&\times
	C^{RM}_{r_{1}m_{1}r_{2}m_{2}}
	C^{RN}_{r_{1}n_{1}r_{2}n_{2}}
	C^{R'''M'''}_{RMr_{3}m_{3}}
	C^{R'''N'''}_{RNr_{3}n_{3}}
	C^{R''M''}_{r_{4}m_{4}R'M'}
	C^{R''N''}_{r_{4}n_{4}R'N'}
	C^{R'M'}_{r_{5}m_{5}r_{6}m_{6}}
	C^{R'N'}_{r_{5}n_{5}r_{6}n_{6}}
	\nonumber\\
	&\times
	\frac{1}{2R''+1}
	\sum_{r_{H}=|R'''-R''|}^{R'''+R''}
	F_{r_{H}}(2H)
	C^{R''M''}_{R'''M'''r_{H}(M''-M''')}
	C^{R''N''}_{R'''N'''r_{H}(M''-M''')}
	.
\end{align}
Note with an external field there is an intermediate tensor of the form
\begin{align}
\label{eq:hmatrix}
    H_{R''' R''}(H) = \frac{1}{2R''+1}
	\sum_{r_{H}=|R'''-R''|}^{R'''+R''}
	F_{r_{H}}(2H)
	C^{R''M''}_{R'''M'''r_{H}(M''-M''')}
	C^{R''N''}_{R'''N'''r_{H}(M''-M''')}
\end{align}
which replaces the Kronecker deltas found in Eq.~\eqref{eq:def_t}.  

Now, with Eq.~\eqref{eq:fwidetilde }, we can write down the necessary impure tensor for the calculation of the magnetization density in Eq.~\eqref{eq:mag},
\begin{align}
	&
	Q_{n;
	x y z z' y' x'
	}
	=
	\sqrt{\prod_{p=1}^{6}F_{r_{p}}( J)}
	\sum_{R=|r_{1}-r_{2}|}^{r_{1}+r_{2}}
	\sum_{R'''=|R-r_{3}|}^{R+r_{3}}
	\sum_{R''=|r_{4}-R'|}^{r_{4}+R'}
	\sum_{R'=|r_{5}-r_{6}|}^{r_{5}+r_{6}}
	\sum_{M,N}
	\sum_{M''',N'''}
	\sum_{M'',N''}
	\sum_{M',N'}
	\nonumber\\
	&\times
	C^{RM}_{r_{1}m_{1}r_{2}m_{2}}
	C^{RN}_{r_{1}n_{1}r_{2}n_{2}}
	C^{R'''M'''}_{RMr_{3}m_{3}}
	C^{R'''N'''}_{RNr_{3}n_{3}}
	C^{R''M''}_{r_{4}m_{4}R'M'}
	C^{R''N''}_{r_{4}n_{4}R'N'}
	C^{R'M'}_{r_{5}m_{5}r_{6}m_{6}}
	C^{R'N'}_{r_{5}n_{5}r_{6}n_{6}}
	\nonumber\\
	&\times
	\frac{1}{2R''+1}
	\sum_{r_{H}=|R'''-R''|}^{R'''+R''}
	\widetilde {F}_{r_{H}}(2H)
	C^{R''M''}_{R'''M'''r_{H}(M''-M''')}
	C^{R''N''}_{R'''N'''r_{H}(M''-M''')}
	.
\end{align}

\subsubsection{Fixed-point tensor analysis}
To understand the behavior of the $O(4)$ fixed-point structure in this model and to check the reliability of our coarse-graining procedure, we use the observable introduced in 
Ref.~\cite{PhysRevB.80.155131}, $X$. This quantity has the property that when computed using fixed-point tensors $T_{ijklmn}$---which are invariant under scaling symmetry: $T \to \Gamma T$---it is unchanged.
In this subsection, we use $X$ to determine the location of $ J_{c}$ i.e., the critical coupling, and to potentially point out the relevant symmetry breaking responsible for the phase transition.  This is possible because the fixed-point structure of the tensor after a large number of iterations can be used to identify the ground state degeneracy of the model. In three dimensions, $X$ is defined as: 
\begin{align}
    X \equiv \frac{(T_{abccba})^{2}}{T_{abccbd}  T_{deffea}},
\end{align}
with the tensor indices arranged as $T_{xyzz'y'x'}$ and summation over repeated indices implied. 
This quantity has the property that when the tensor is a direct sum of $n$ dimension-one tensors, $X=n$, and when it is a simple dimension-one tensor, $X=1$ (in the disordered phase at high temperatures or small $J$). We can use this property to identify when the model undergoes a change in its ground state degeneracy corresponding to a phase transition.

\subsection{Definition of initial tensors for tensor renormalization groups}
\label{sec:tensor-decomposition}
\begin{figure}
    \centering
    \includegraphics[width=16cm]
    {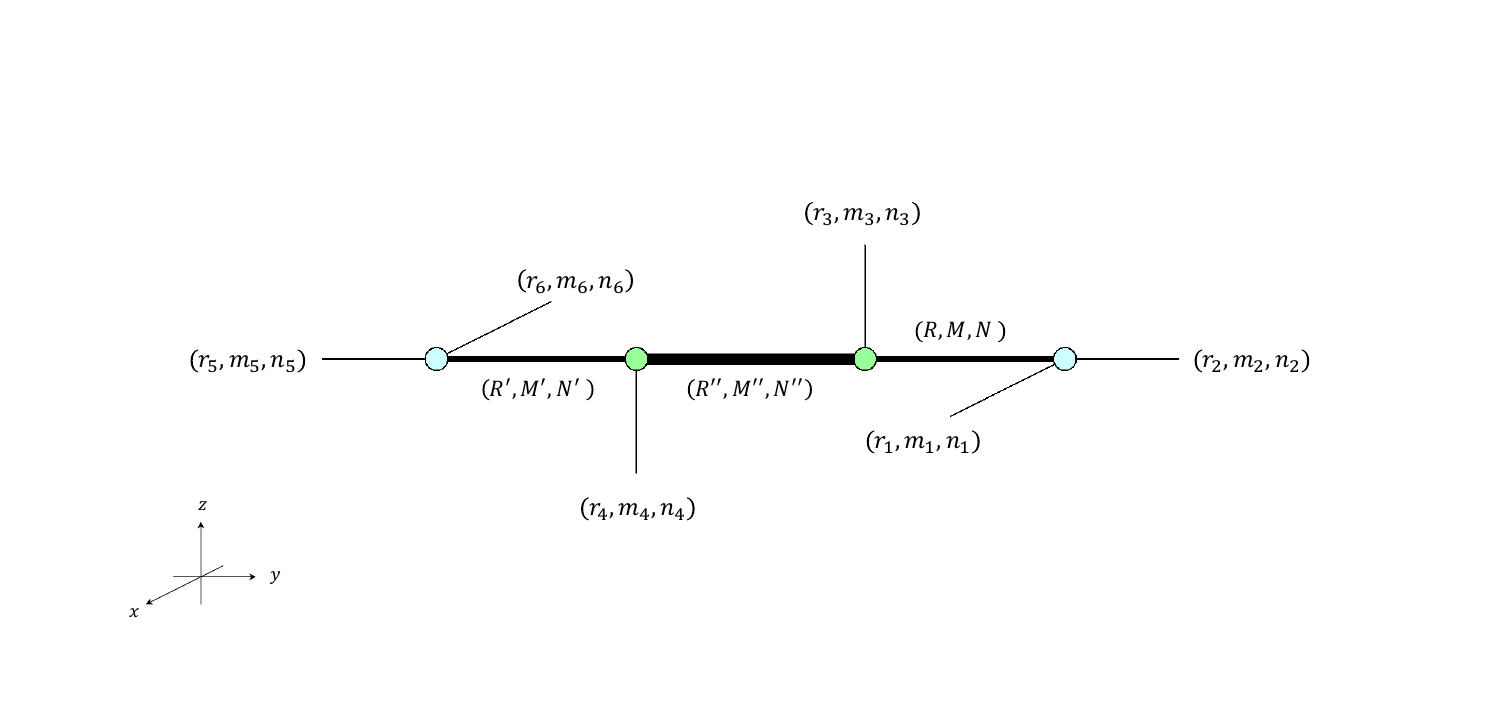}
    \caption{
    Initial tensors for the tTRG.
    }
    \label{fig:triad}
\end{figure}

\begin{figure}
    \centering
    \includegraphics[width=16cm]%
    {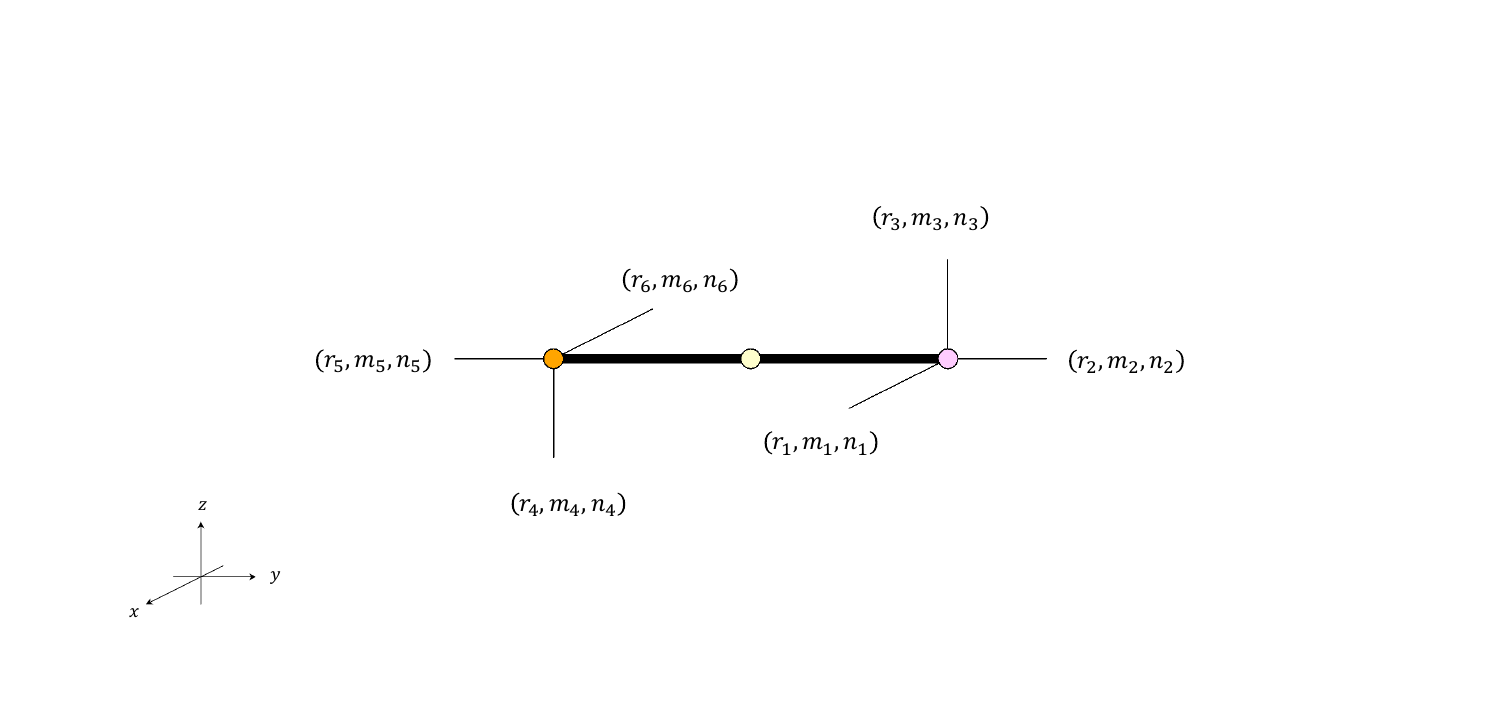}
    \caption{
    Initial tensors for the ATRG in Eq.~\eqref{eq:svd_t}.
    Four-leg tensors denote unitary matrices and the two-leg tensor shows the singular value.
    In the practical computation, the size of the internal line is truncated by the bond dimension $D$ in the ATRG.
    }
    \label{fig:atrg}
\end{figure}

The tensor network contraction formally given in Eq.~\eqref{eq:tn_z} cannot be carried out exactly, except in the most simple of cases.  Instead, we use the tTRG and ATRG methods to approximate it.  To use these methods, the primary tensor, $T$, must be decomposed into an appropriate form for each specific algorithm.  

The philosophy of the tTRG is to re-write the initial tensor into a contraction of four smaller tensors (in three dimensions).  For a generic six-indexed tensor in three dimensions, this takes the form,
\begin{equation} 
\label{eq:triad-form}
T_{x y z z' y' x'}=\sum_{a, b, c} A_{x y a} B_{a z b} C_{b z' c} D_{c y' x'}.
\end{equation} 
The principal chiral model naturally takes the form of a triad decomposition.  The right-hand-side of Eq.~\eqref{eq:triad-form} is in the same form as Eq.~\eqref{eq:def_t}, with an identification of $x=(r_{1} m_{1} n_{1})$, $x' = (r_{6} m_{6} n_{6})$ etc. which gives for the triad tensors,
\begin{align}
& A_{x y a} = \sqrt{F_{r_{1}}( J) F_{r_{2}}( J)} {C}_{r_{1} m_{1} r_{2} m_{2}}^{R M} {C}_{r_{1} n_{1} r_{2} n_{2}}^{R N} \\
& B_{a z b} = \frac{1}{\sqrt{d_{R''}}} \sqrt{F_{r_{3}}( J)} C^{R''M''}_{RMr_{3}m_{3}} C^{R''N''}_{RNr_{3}n_{3}} \\
& C_{b z' c} = \frac{1}{\sqrt{d_{R''}}} \sqrt{F_{r_{4}}( J)} C^{R''M''}_{r_{4}m_{4}R'M'}
	C^{R''N''}_{r_{4}n_{4}R'N'} \\
& D_{c y' x'} = \sqrt{F_{r_{5}}( J) F_{r_{6}}( J)} C^{R'M'}_{r_{5}m_{5}r_{6}m_{6}}
	C^{R'N'}_{r_{5}n_{5}r_{6}n_{6}},
\end{align}
where the index structure makes it clear when we are referring to the Clebsch-Gordan coefficients, and when we are referring to the $C$ tensor.
This triad formulation is shown in Fig.~\ref{fig:triad}, with the $A$, $B$, $C$, and $D$ tensors appearing from right to left.

The impure tensors from Eqs.~\eqref{eq:Rimp} and~\eqref{eq:Simp} can be immediately used in the triad formulation as well.  They modify the, say, $B$ and $C$ tensors on two adjacent sites, say, $n$ and $n+\hat{z}$, respectively,
\begin{align}
\widetilde {B}_{a z b}(n) &= \sqrt{\frac{\widetilde {F}_{r_{3}}( J)}{F_{r_{3}}( J)}} B_{a z b}(n) \\ \nonumber
&=\frac{1}{\sqrt{d_{R''}}} \sqrt{\widetilde {F}_{r_{3}}( J)} C^{R''M''}_{RMr_{3}m_{3}} C^{R''N''}_{RNr_{3}n_{3}} \\
\widetilde {C}_{b z' c}(n+\hat{z}) &= \sqrt{\frac{\widetilde {F}_{r_{4}}( J)}{F_{r_{4}}( J)}} C_{b z' c}(n+\hat{z}) \\ \nonumber
&=\frac{1}{\sqrt{d_{R''}}} \sqrt{\widetilde {F}_{r_{4}}( J)} C^{R''M''}_{r_{4}m_{4}R'M'}
	C^{R''N''}_{r_{4}n_{4}R'N'}.
\end{align}
For the case of an external field, the $H$ matrix in Eq.~\eqref{eq:hmatrix} naturally fits between the $B$ and $C$ tensors.

For the ATRG,
we convert $T$ into the canonical form to evaluate Eq.~\eqref{eq:tn_z}.  This decomposition can be seen in Fig.~\ref{fig:atrg}.
The canonical form we need is
\begin{align}
\label{eq:svd_t}
	&T_{
	x y z z' y' x'
	}
	=
	\sum_{\gamma}
	U_{
	x y z
	\gamma
	}
	\sigma_{\gamma}
	V^{*}_{
	z' y' x'
	\gamma
	}
	,
\end{align}
which is nothing but the SVD of $T$.  This canonical form can be constructed without using the full six-indexed tensor by using the natural triad structure of the tensor described above.
In the practical computation, we introduce the low-rank approximation in Eq.~\eqref{eq:svd_t} to decimate the smaller singular values up to the bond dimension.

\section{Results}
\label{sec:results}
\begin{figure*}
    \centering
    \includegraphics[width=17.2cm]{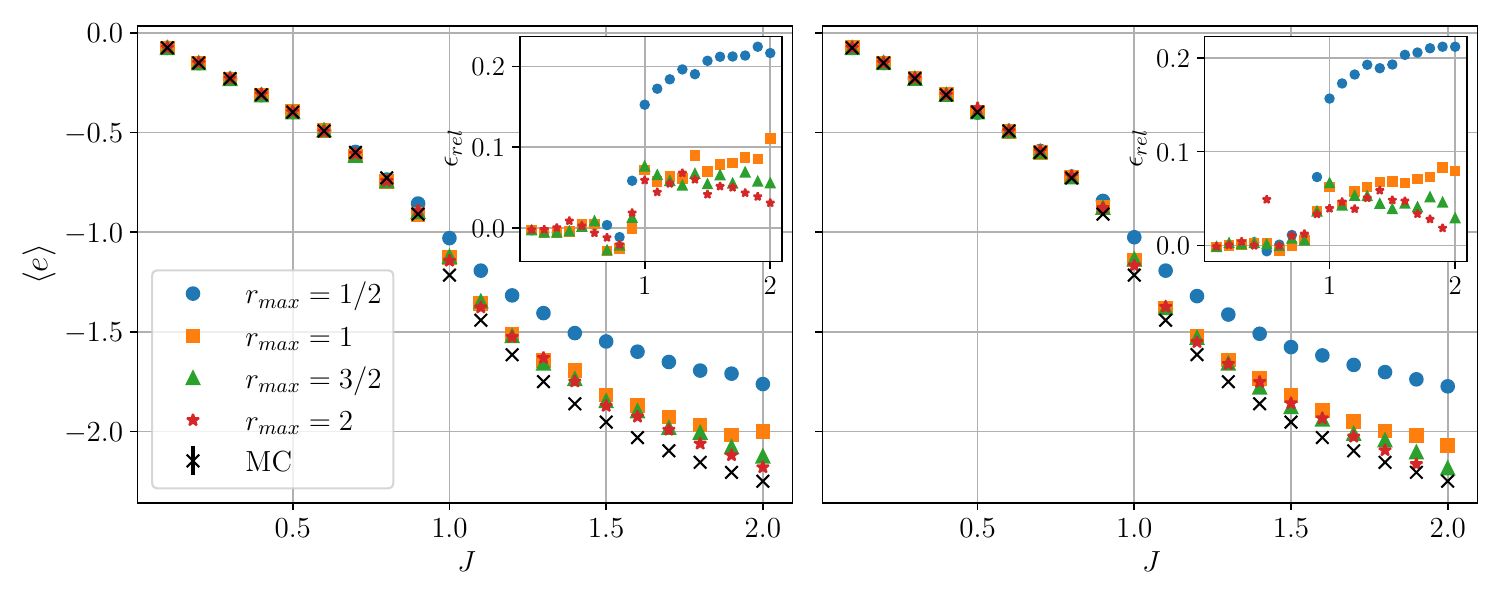}
    \caption{The average internal energy plotted alongside the MC computation using the tTRG (left) and ATRG (right). In both tensor computations $D=50$ for the three lowest irreps., and $D=70$ for $r_{\rm max}=2$, and $L=16$.}
    \label{fig:act-mc-comp}
\end{figure*}
\begin{figure}
    \centering
    \includegraphics[width=8.6cm]{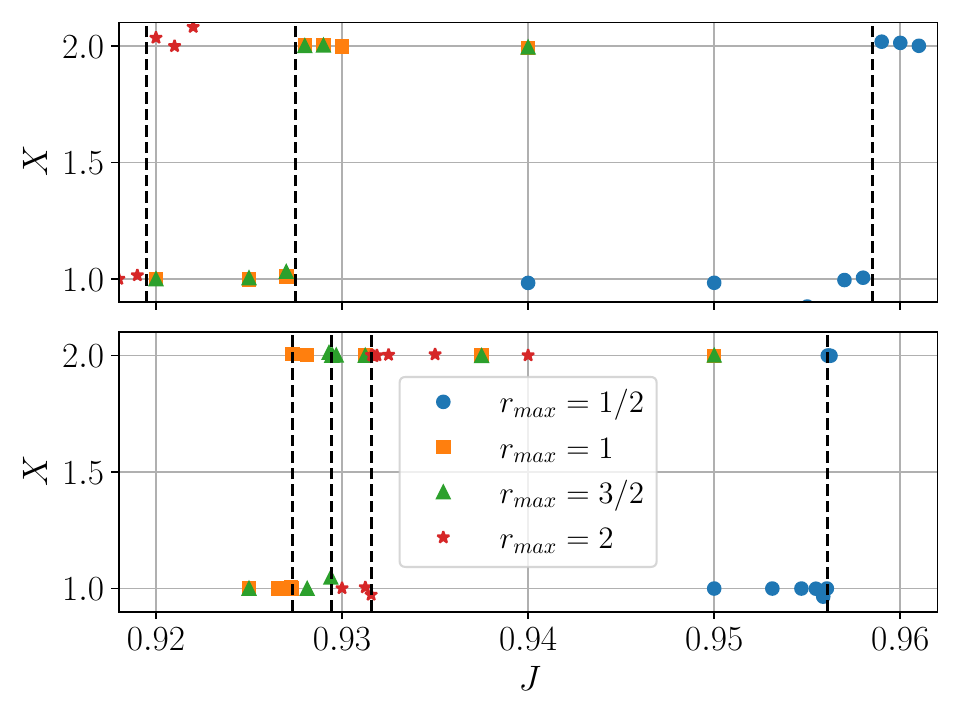}
    \caption{The $X$ quantity using the tTRG (top) and ATRG (bottom).Both methods used $D=50$, and $L=1024$.  The vertical dashed lines indicate the estimated values for the various $J_{c}$. The correct MC value is $J_{c} = 0.936(1)$.      }
    \label{fig:xobs}
\end{figure}
\begin{figure}
    \centering
    \includegraphics[width=8.6cm]{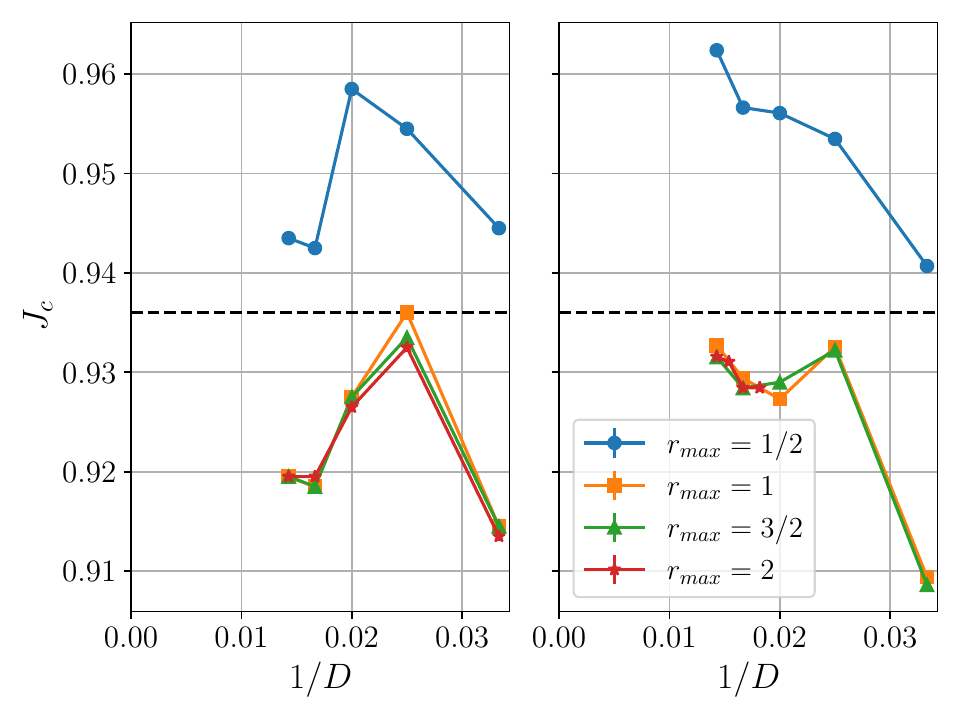}
    \caption{The critical coupling $J_{c}$ as determined using $X$ plotted versus $1/D$. The result from MC~\cite{Kanaya:1994qe} is shown with a dashed line and an uncertainty that is not visible in the plot.}
    \label{fig:jc-converge}
\end{figure}

With the tensor constructions defined in Sec.~\ref{sec:tensor-decomposition}, we can compute quantities using both the tTRG, and the ATRG, and compare.  For this comparison, we consider $\braket{e}$, $X$, and $\braket{m}$, and we make the comparison at fixed bond dimension.

The computation of the internal energy can be seen in Fig.~\ref{fig:act-mc-comp}.  This is done using the impure tensor.  Here we not only compare the tTRG and the ATRG but also include MC results to provide a cross-check.  The relative error,
\begin{align}
    \epsilon_{\text{rel}} = \frac{\braket{e}_{\text{TRG}} - \braket{e}_{\text{MC}}}{\braket{e}_{\text{MC}}},
\end{align}
between the tensor calculations and the MC are shown in the inset plot.  We find in the ``strong-coupling/high-temperature'' regime that the tensor results agree well with the MC results.  In the ``weak-coupling/low-temperature'' regime we find poorer accuracy, however, as $r_{\rm max}$ is increased, the tensor results move towards the MC data.

In Fig.~\ref{fig:xobs} we see the comparison between the tTRG and the ATRG in computing $X$.  Previous studies~\cite{PhysRevB.80.155131,Wang_2014,Chen:2017ums,Akiyama:2019xzy,Jha:2022pgy} focused on computing $X$ in the presence of a discrete global symmetry. Its use as a tool to identify phase transitions for continuous global symmetries is untested. Both tensor methods see a ``jump'' in $X$ around some $J_{c}$, indicating a potential phase transition.  We use this jump to identify $J_{c}$ for various $r_{\rm max}$ values, which we use in later computations of the magnetization.  The location of the jumps as a function of $r_{\rm max}$ does not completely agree between the two methods, although they are roughly consistent.  However, the tTRG value for $J_{c}$ with $r_{\rm max}=1$ and $r_{\rm max}=3/2$ are the same at this resolution, unlike the ATRG which has two different values for these two $r_{\rm max}$ cutoffs.  The tTRG result for $r_{\rm max}=2$ actually worsens, in contrast to the ATRG.

The values of $J_{c}$ extracted using this method can be seen plotted against $D$ in Fig.~\ref{fig:jc-converge}. The result obtained using MC methods shows $J_{c} = 0.936(1)$ \cite{Kanaya:1994qe} which is closest to the ATRG result with the highest truncation. It seems that with these results, ATRG does slightly better than tTRG in obtaining a better approximation.
It is worth noting that the jump from $X = 1$ to $X = 2$ is reminiscent of a similar jump associated with the Ising universality class.  However, as found in MC, as well as in this study, the critical exponents found for the phase transition in this model are not consistent with the Ising type as we will see below.  

\begin{figure}
    \centering
    \includegraphics[width=8.6cm]{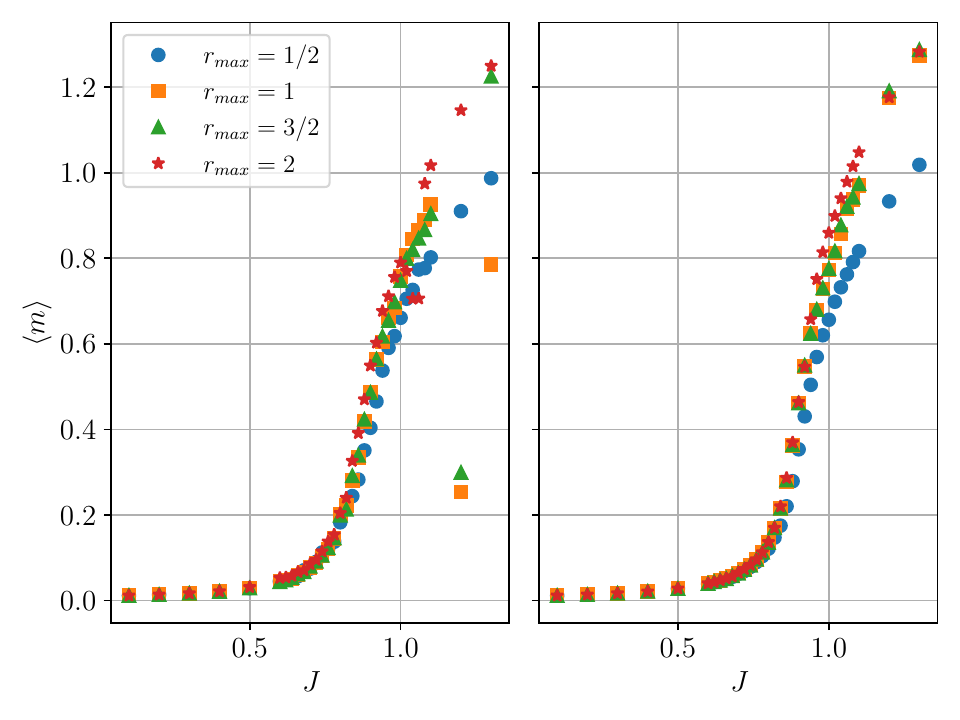}
    \caption{Magnetization with $H=0.01$ for tTRG (left)
    and ATRG (right). We use $D=50$ for the three smallest irreps. and $D=70$ for $r_{\rm max}=2$.  The lattice volume is $1024^3$ in all cases.}
    \label{fig:mag-compare}
\end{figure}

The final quantity we consider is the magnetization.  This quantity is calculated in the presence of an external field using an impure tensor.  The results of this calculation can be seen in Fig.~\ref{fig:mag-compare} for $H=0.01$.  We see somewhat similar results between the tTRG and ATRG, however, at larger $J$ values, the tTRG possesses noise which can be radical for some values of $J$.  We find this effect is both $r_{\rm max}$ and $D$ dependent.  
One can also see that the tTRG result is systematically slightly lower than the ATRG result, especially in the large $J$ regime, indicating the tTRG may be missing a fraction of the true critical behavior.

\begin{figure}[htbp]
    \centering
    \subfigure[$r_{\text{max}}=1/2$]{
    \includegraphics[width=8.6cm]{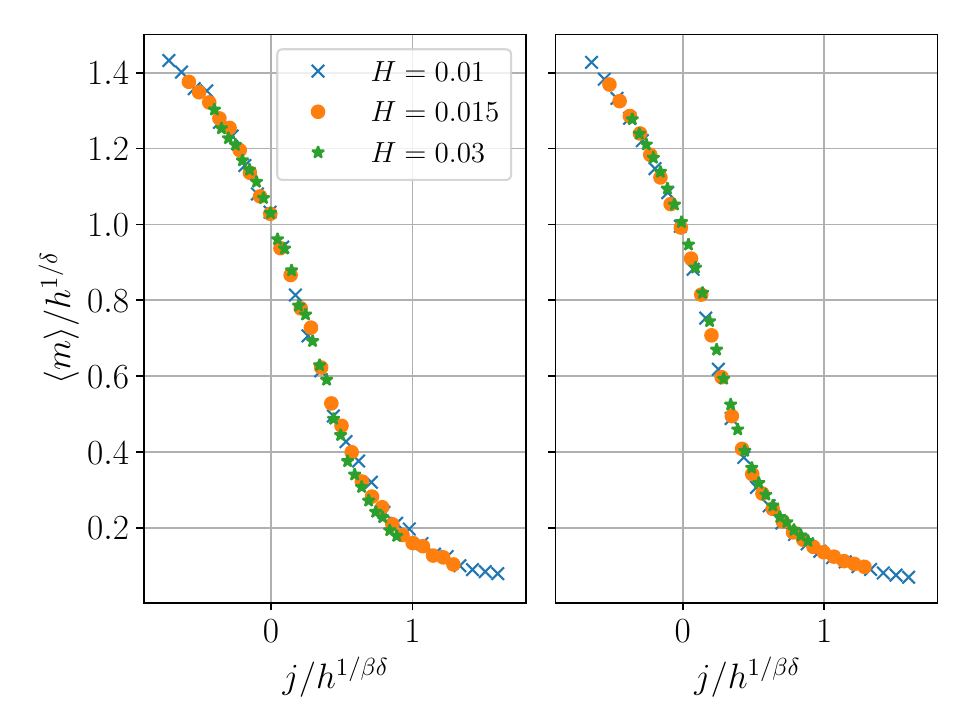}
    }
    \subfigure[$r_{\text{max}}=1$]{
    \includegraphics[width=8.6cm]{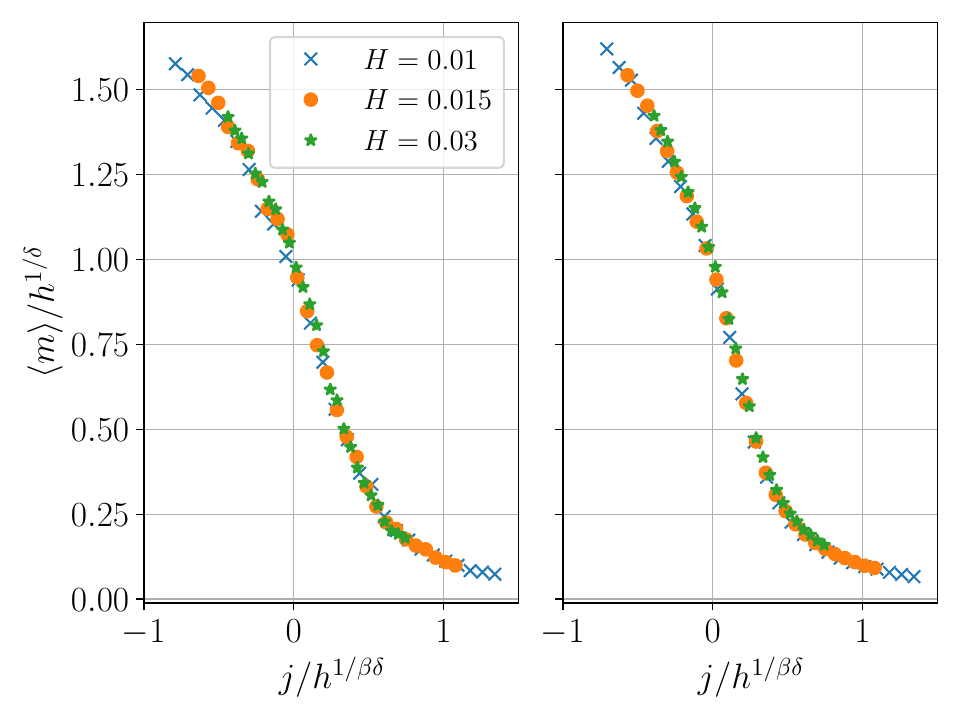}
    }
    \subfigure[$r_{\text{max}}=3/2$]{
    \includegraphics[width=8.6cm]{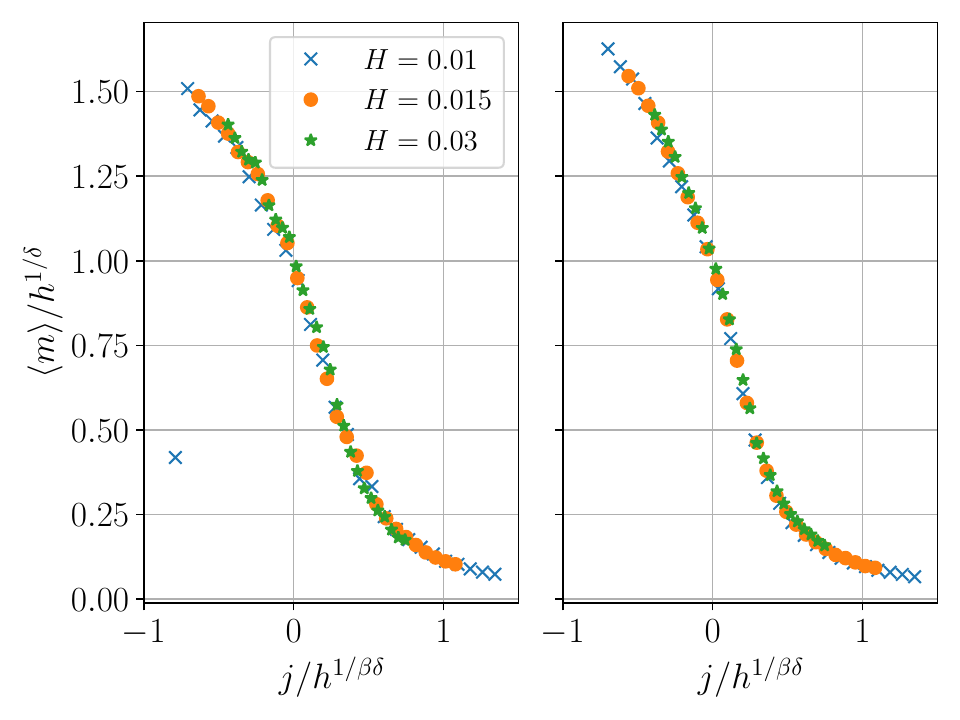}
    }
    \subfigure[$r_{\text{max}}=2$]{
    \includegraphics[width=8.6cm]{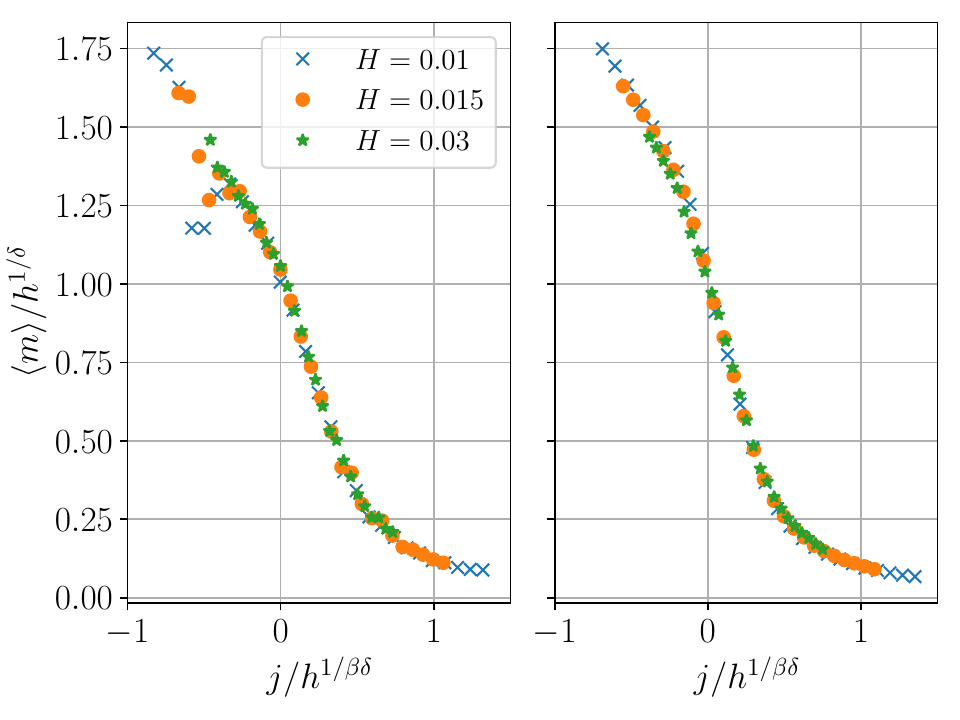}
    }
    \caption{The magnetization collapse for various truncations. The left panel shows the results using tTRG, and the right panel shows ATRG. $r_{\text{max}}=1/2, 1, 3/2$ were calculated with $D=50$, and $r_{\text{max}}=2$ was done with $D=70$. The volume is $1024^{3}$.
    }
    \label{fig:mag-collapse} 
\end{figure}

To address this point, we attempt to collapse the magnetization data under the assumption of the existence of a critical point, which was carried out using MC and non-perturbative renormalization group method for this model in Refs.~\cite{PhysRevD.55.362,Braun:2007td,Engels:2011km}.  
We define the following variables,
\begin{align}
    j \equiv \frac{J_{c} - J}{J_{c}},
    \label{eq:smallj_def}
\end{align}
and
\begin{align}
    h \equiv \frac{H}{H_{0}},
     \label{eq:smallh_def}
\end{align}
which can be used in the following scaling functional form
\begin{align}
    \frac{\braket{m}}{h^{1/\delta}} = f(j / h^{1/\beta \delta})
\end{align}
that is valid in the regime of critical behavior.  
Figure~\ref{fig:mag-collapse} shows the resulting collapsed data for four different truncations.  Here $H_{0}$ is chosen such that $f(0) = 1$, and $\beta$ and $\delta$ are the usual critical exponents.  We use the values from Ref.~\cite{Kanaya:1994qe} of $\beta = 0.3836(46)$ and $\delta = 4.851(22)$.  The respective values of $J_{c}$ are determined using the results from the $X$ calculation, for both the tTRG and ATRG collapses.  We find the TRG results are consistent with the critical exponents obtained using MC using both methods, indicating that while the tTRG estimate for $J_{c}$ is worse than the ATRG estimate, the underlying critical behavior captured by it for the magnetic exponents is still consistent with the literature.\footnote{
One can fit $\alpha$ directly from the internal energy using the following expression~\cite{Xie:2012mjn},
\begin{align}
  \label{eq:alpha-fit}
  \braket{e} = A + B|J - J_{c}| + C|J-J_{c}|^{1-\alpha},
\end{align}
where $A$, $B$, and $C$ are fit parameters. Using this method, both the tTRG and the ATRG finds $\alpha$ that is consistently negative~\cite{Akiyama:2023fyk}. However, obtaining a precise value following this way is difficult and we leave this for future work.
}

\section{Conclusions}
We carried out a systematic study of the three-dimensional $SU(2)$ principal chiral model which belongs to the same universality class as the $O(4)$ nonlinear sigma model. This model is useful in understanding chiral symmetry in QCD in a simplified setting. 
Since our tensor network formulation is based on character expansion, the $SU(2)$ symmetry is explicitly present even with a finite $r_{\rm max}$ truncation.
We used different tensor network algorithms and computed the critical coupling, average action, magnetization, and fixed point observables.
In addition, we also compute the RG collapse plots for magnetization which has previously not been attempted using the higher-dimensional TRG approach. Our results show the efficiency of tensor methods even in three Euclidean dimensions to reproduce the expected critical exponents. 
Our results are also the first direct comparison of observables, calculated using two different methods (ATRG and tTRG), in a model with a continuous non-Abelian symmetry group
in three dimensions.

Generally, we find the ATRG is more accurate and less noisy with respect to the results from the literature; however, the tTRG does correctly capture the critical behavior for the exponents calculated here.  Moreover, we note that the tTRG was compared to the ATRG using the same $D$ in all calculations, but that the tTRG possesses superior scaling in computational cost which was not taken advantage of in these calculations.  Future studies using the full strength of the tTRG could demonstrate improved accuracy.
It would be interesting to improve the algorithms used in this paper such that those computations can be pursued. We leave this improvement and study of gauge theories in three dimensions for future work. 

\begin{acknowledgments}
We thank Jacques Bloch for sharing the Monte Carlo data and Katsumasa Nakayama for sharing several data presented in Ref.~\cite{Kadoh:2019kqk}.

SA acknowledges the support from the Endowed Project for Quantum Software Research and Education, the University of Tokyo~\cite{qsw}, JSPS KAKENHI Grant Number JP23K13096, and the Center of Innovations for Sustainable Quantum AI (JST Grant Number JPMJPF2221).
The ATRG calculation for the present work was carried out with ohtaka provided by the Institute for Solid State Physics, the University of Tokyo, and the use of SQUID at the Cybermedia Center, Osaka University (Project ID: hp240012).

RGJ is supported by the U.S. Department of Energy, Office of Science, National Quantum Information Science Research Centers, Co-design Center for Quantum Advantage (C2QA) under contract number DE-SC0012704 and by the U.S. Department of Energy, Office of Science, Office of Nuclear Physics under contract number DE-AC05-06OR23177.

This manuscript has been co-authored by an employee of Fermi Research Alliance, LLC under Contract No. DE-AC02-07CH11359 with the U.S. Department of Energy, Office of Science, Office of High Energy Physics.  This work is supported by the Department of Energy through the Fermilab Theory QuantiSED program in the area of ``Intersections of QIS and Theoretical Particle Physics''.

We thank the Institute for Nuclear Theory at the University of Washington for its kind hospitality and stimulating research environment. This research was supported in part by the INT's U.S. Department of Energy grant No. DE-FG02-00ER41132.
\end{acknowledgments}

% \bibliographystyle{utphys.bst}
% \bibliography{refs1.bib}

\end{document}